\documentclass[12pt]{article}
\usepackage{graphicx, natbib}

\newcommand\pubnumber{Article 19 in eConf C1304143}
\newcommand\pubdate{\today}

\def\nice{$^1$ARTEMIS (CNRS/UNS/OCA);  
$^2$Osservatorio Astronomico di Roma, OAR-INAF; $^3$Universit\'e de Toulouse; UPS-OMP; IRAP; 
$^4$CNRS; IRAP; $^5$Aix Marseille Universit\'e, CNRS, LAM; $^6$CNRS; Observatoire de Haute-Provence;
$^7$University of Western Australia, School of Physics; $^8$ASI Science Data Center; OAR-INAF;
$^9$Mullard Space Science Laboratory; UCL; $^{10}$IAPS, INAF}

\def\Title#1{\begin{center} {\Large #1 } \end{center}}
\def\Author#1{\begin{center}{ \sc #1} \end{center}}
\def\Address#1{\begin{center}{ \it #1} \end{center}}

\newcommand\pubblock{\rightline{\begin{tabular}{l} \pubnumber\\
         \pubdate  \end{tabular}}}
\newenvironment{Abstract}{\begin{quotation}  }{\end{quotation}}
\newenvironment{Presented}{\begin{quotation} \begin{center} 
             PRESENTED AT\end{center}\bigskip 
      \begin{center}\begin{large}}{\end{large}\end{center} \end{quotation}}
\def\Acknowledgements{\bigskip  \bigskip \begin{center} \begin{large}
             \bf ACKNOWLEDGEMENTS \end{large}\end{center}}
\def\grb{GRB 111209A}
\def\apj{ApJ}
\def\mnras{MNRAS}

\def\beq{\begin{equation}}
\def\eeq#1{\label{#1}\end{equation}}
\def\eeqn{\end{equation}}

\def\beqa{\begin{eqnarray}}
\def\eeqa#1{\label{#1}\end{eqnarray}}
\def\eeqan{\end{eqnarray}}

\let\bar=\overbar

\def\Dslash{\not{\hbox{\kern-4pt $D$}}}
\def\dslash{\not{\hbox{\kern-2pt $\del$}}}

\def\msb{{\bar{\ssstyle M \kern -1pt S}}}

\begin{document}
\begin{titlepage}
\pubblock

\vfill
\Title{The diversity of progenitors and emission mechanisms for ultra-long bursts}
\vfill
\Author{ B. Gendre$^1$, G. Stratta$^2$, J.L. Atteia$^{3,4}$, S. Basa$^5$, M. Bo\"er$^{1,6}$, D.M. Coward$^7$, S. Cutini$^8$, V. D'Elia$^8$, E.J. Howell$^7$, A. Klotz$^{3,4,6}$, S. Oates$^9$, M. De Pasquale$^9$, L. Piro$^{10}$}
\Address{\nice}
\vfill
\begin{Abstract}
GRB 111209A is the longest ever recorded burst. This burst was detected by Swift and Konus-Wind, and we obtained TOO time from XMM-Newton as well as prompt data from TAROT. We made a common reduction using data from these instruments together with other ones. This allows for the first time a precise study at high signal-to-noise ratio of the prompt to afterglow transition. We show that several mechanisms are responsible of this phase. In its prompt phase, we show that its duration is longer than 20 000 seconds. This, combined with the fact that the burst fluence is among the top 5\% of what is observed for other events, makes this event extremely energetic. We discuss the possible progenitors that could explain the extreme duration properties of this burst as well as its spectral properties. We present evidences that this burst belong to a new, previously unidentified, class of GRBs. The most probable progenitor of this new class is a low metalicity blue super-giant star. We show that selection effects could prevent the detection of other bursts at larger redshift and conclude that this kind of event is intrinsically rare in the local Universe. The afterglow presents similar features to other normal long GRBs and a late rebrightening in the optical wavelengths, as observed in other long GRBs. A broad band SED from radio to X-rays at late times does not show significant deviations from the expected standard fireball afterglow synchrotron emission. 
\end{Abstract}
\vfill
\begin{Presented}
GRB 2013, the Seventh Huntsville Gamma-Ray Burst Symposium \\
Nashville, Tennessee, 14--18 April 2013
\end{Presented}
\vfill
\end{titlepage}
\def\thefootnote{\fnsymbol{footnote}}
\setcounter{footnote}{0}
%



\section{Introduction}

Gamma-ray bursts (GRBs) are detected as brief flashes of high-energy photons typically lasting some tens of seconds \citep[see][for a review]{mes06}. We here present the summary of an analysis of the longest burst ever observed, \grb, which had a duration of 25000 seconds. We used data obtained in $\gamma$-ray, X-ray, optical and radio. Details can be found in \citet{gen13} (hereafter Paper I) and \citet{str13} (Paper II). In the following, we assume flat $\Lambda$ CDM Universe with $H_0 = 71$ km s$^{-1}$ Mpc$^{-1}$, $\Omega_m = 0.27$, and quote all errors at the 90 \% confidence level. 

\section{GRB 111209A}
\label{sec_obs}

\grb was discovered by the Swift satellite at $T_0$ = 2011:12:09-07:12:08 UT \citep{hov11}. However, the burst started about 5400 seconds before $T_0$, as shown in the ground data analysis of the Konus-Wind instrument. Swift did not trigger at the start of the event as the burst was not in the field of view of the BAT instrument. The gamma-ray signal was observed up to about $T_0$ + 10000 s \citep{gol11}.

Swift/XRT observations started 425 s after the BAT trigger \citep{hov11} revealing a bright afterglow, observed also by Swift/UVOT in the optical-UV bands. The afterglow was also clearly detected by ground based instruments; for example the TAROT-La Silla \citep{klo11}, the GROND robotic telescopes \citep{kan11}, Gemini and VLT telescopes \citep{kan11, lev13}. The Australia Telescope Compact Array (ATCA) observed this event in two bands, 5.5 and 9 GHz, at the mean observing time of $T_0+446.4$ ks \citep{han11}. In addition, we activated a Target of Opportunity observation with XMM-Newton, between $T_0 + 56,664$ s and $T_0 + 108,160$ s. This period covered the end of the prompt phase seen in X-ray, a subsequent plateau phase, and the start of the normal afterglow decay \citep{gen11}. We refer the reader to Papers I and II for the detail of data reduction.

\section{Prompt phase}

We extracted several Spectral Energy Distributions (SEDs) at 650-850 s, 1-1.2 ks, and 7-8 ks after $T_0$. We found that the high energy spectrum (above $\sim 1$ keV) is consistent with a single power law, while the optical data imply a non negligible absorption by dust in the host galaxy, using a Milky Way dust model ($A_V \sim 0.9 - 1.5$). Such a large value of $A_V$ has already been invoked to model the prompt emission of other bursts \citep{gor12}. In addition, the data need the addition of a black body component in the soft X - far UV energy range, accounting for about 0.01 \% of the total flux in the 0.5-10 keV band. This component is not detected at later times.

Between $\sim T_0+400$ s and $\sim T_0+4000$ s the prompt emission was observed from optical wavelengths to gamma-rays. A flare is observed at $\sim T_0+2$ ks and clearly visible at all wavelengths, with a peak epoch at $T_0 + (2460\pm50)$ s in optical, and $T_0 + (2050\pm10)$ s in $\gamma$-ray. The rest-frame delay of the R band peak epoch is $\sim245$ s. This is far more larger that what was reported for other events (usually a few seconds, see e.g. \citep{Klotz2009}. A possible explanation is that the optical counterpart of the gamma-ray flare is generated in a different region of the ejecta, more distant from the one producing the $\gamma$-rays.

\section{Steep decay and the plateau phase}

The XMM-Newton monitoring period covered the transition from the steep decay phase to the ``normal" decay phase, for an interval of 53 000 seconds. A double power law model best fit the data. The spectral fit clearly indicates that this second component is not dominating, and provides a small contribution to the total observed flux. Some bursts observed by Fermi shows the presence of a very hard power law spectral component in addition to the typical prompt emission spectrum. This extra-component decays once the afterglow has started \citep[e.g.][]{Abdo2009}. One may thus consider the extra-component seen in GRB 111209A as the "soft tail" of this hard power law seen by the Fermi/LAT. Among the possible interpretations, \citet{Zhang2011} proposed that it may originate from the Compton-up scattered emission of a simultaneous thermal emission in the MeV energy range. Although we detect a thermal emission for GRB 111209A, it was in the soft X-ray energy range and not simultaneous to the power law extra-component. An alternative scenario suggests this component to be emitted from another site using the classical emission mechanism for GRBs \citep{Zhang2011}. 

\section{The afterglow}

Nearly simultaneous radio, optical and X-rays data are available for this burst at late time. The X-ray spectral and temporal indices are consistent with the synchrotron model expectations for $\nu_X>\nu_c$. The optical to X-ray data are best fit by a broken power law model with a negligible amount rest frame visual dust extinction. We note that this is different from the prompt phase, where more dust was needed, and can be explained by dust destruction in the vicinity of the progenitor \citep[e.g.][]{Waxman2000}. 

Radio data alone provide excellent agreement with the expected 1/3 spectral slope if $\nu_{radio}$ were below the spectral peak frequency. Following \cite{Panaitescu2000}, we could fit the multi-band SED  for $\nu_{radio}<\nu_m<\nu_{opt}<\nu_c<\nu_X$ with an environment density $n=0.07$ cm$^{-3}$, a fraction of the total energy transferred to the magnetic field of $\epsilon_B=0.0003$ and to the swept-up electrons of $\epsilon_e=0.03$, $p=2.6$, $\eta=0.1$ and a collimation factor of 0.08 (i.e. a half opening angle of 23 degrees). We could not find any obvious solution assuming a wind environment.

\section{Duration of the prompt emission}

We estimated the total duration of this event to be at least 25,000 seconds. We found no other ultra-long candidate in the BATSE 4B catalog \citep{pac99} or for Swift and Fermi. We note however that four BATSE events might represent in fact one single GRB \citep{con98}. We estimated the rate for ultra long GRBs to be $\sim 6 \times 10^{-5}$ Gpc$^{-3}$ yr$^{-1}$ in the local Universe, significantly lower than that of the normal long GRB \citep[see][]{how13}.

It is possible that selection effects make only the brightest part of ultra-long burst light curves detectable, thus reducing their intrinsic durations and making ultra-long GRBs indistinguishable from normal bursts. We found that this event can be detected only within a limiting redshift of z = 1.4. However, for this distance range, it appears every time as an ultra-long GRB.

\section{Energetics}

\begin{figure}
\includegraphics[width=9cm]{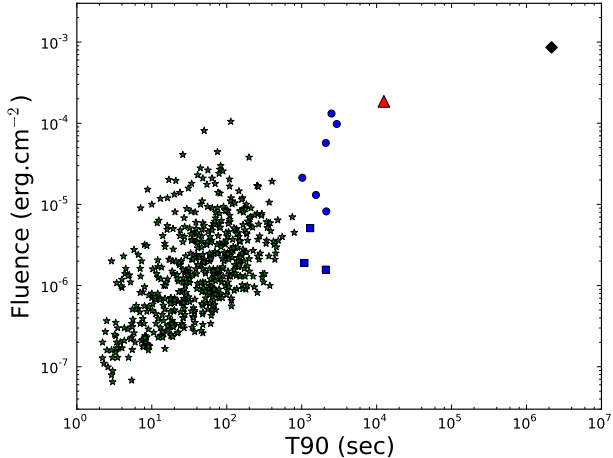}
\caption{Position of GRB 111209A (red triangle) in the GRB Fluence-Duration plane. The green stars are long Swift bursts. Blue circles and squares are the recorded super long GRBs, and the black dot is J1644+57. Blue squares represent high-energy transients attributed to SN shock breakout. Note that all measurements are converted to the Swift band (15-150 keV) for comparison with the Swift catalog. Figure extracted from Paper I. \label{fig3}}
\end{figure}

We retrieved from the literature the prompt emission information available for all bursts with a duration larger than 1000 seconds. The distribution of these bursts in the duration-fluence plane is displayed in Fig. \ref{fig3}, together with the tidal disruption event J1644+57. \grb~appears as an outlier, however linked to other very long events. We note that supernova shock breakouts have a much smaller fluence, weakening a possible link between this kind of progenitor and \grb. Correcting for distance effects, this burst also appears to be more energetic than the others in the prompt phase. This is the contrary in the afterglow phase, where its afterglow appears slightly less luminous than normal ones. It can be interpreted as the fact that most of the energy is released in internal shocks, few remaining for the forward shock.

\section{Nature of this event}

The first hypothesis is that the exceptional duration of \grb~is due to a supernova shock breakout \citep{cam06, sta11}. As we do not observe a strong thermal component, contrary to these two bursts, we can discard this hypothesis.

Magnetars could account for long GRB properties \citep[e.g.][]{met10}. However, the magnetar models need unphysical parameter values to reproduce the observations. We can discard this hypothesis too.

A Swift J1644+57 like event is also not very convincing : the light curve should decay as $t^{-5/3}$ \citep{lod09}, which is not observed. 

The main difficulty in explaining the nature of the progenitor of \grb~is its duration. \citet{woo12} investigated several scenarios to provide a long duration event with an energy budget typical of a long GRB. They proposed a single super-giant stars with less than 10\% solar metallicity as progenitor. For a $10^4$ s duration, one needs a blue supergiant in that case. However, spectroscopic analysis have shown that the host galaxy metallicity is sub-solar, but not especially low \citep[about $0.35 Z_{\odot}$][]{lev13}. Because for short life-timescale massive stars, the host metallicity likely reflects the one of the star, the metalicity is too high to support this hypothesis. A possible solution may be that the blue supergiant was formed through a binary system channel \citep[e.g.][]{Podsiadlowski1992,Fryer2005,Podsiadlowski2010}. In such a case the low metalicty condition is relaxed in the models. We thus propose a blue supergiant star as the progenitor of this event.

\section{Conclusion}
\label{sec_conclu}

We have presented multi-wavelength observations of \grb. This event is a very rare kind of GRB, with an ultra-long duration. Its properties allowed several findings :
\begin{itemize}
\item a very large delay in the prompt phase between the optical and high energy emission, possibly linked to two different emission sites
\item a very hard component, compatible with the one seen by Fermi in several short GRBs
\item a large amount of dust during the prompt phase, that disappears during the afterglow, maybe signing some dust destruction by the GRB itself
\item a very unusual blue super-giant progenitor, possibly formed through the merging of a binary
\item the derivation of the parameters of the fireball, from radio to X-ray
\end{itemize}

The impressive quality of X-ray data can allow further researches for features in the prompt spectrum.

\Acknowledgements

This work has been financially supported by ASI grant I/009/10/0 and by the PNHE. 
D.M. Coward is supported by an Australian Research Council Future Fellowship. TAROT has been built with the support of the CNRS-INSU. We thank the technical support of the XMM-Newton staff, in particular N. Loiseau and N. Schartel.

\end{document}